\documentclass{article}

\usepackage{PRIMEarxiv}

\usepackage[utf8]{inputenc} 
\usepackage[T1]{fontenc}    
\usepackage{hyperref}       
\usepackage{url}            
\usepackage{booktabs}       
\usepackage{amsfonts}       
\usepackage{nicefrac}       
\usepackage{microtype}      
\usepackage{lipsum}
\usepackage{fancyhdr}       
\usepackage{graphicx}       
\graphicspath{{media/}}     
\usepackage{multirow}
\usepackage{newunicodechar}
\usepackage{textgreek} 

\title{Short-Term Gains, Long-Term Gaps: The Impact of GenAI and Search Technologies on Retention
}

\author{
  Mahir Akgun \\
  College of Information Sciences and Technology \\
  Penn State University \\
  University Park, PA, USA\\
  \texttt{makgunl@psu.edu} \\
   \And
  Sacip Toker \\
  Information Systems Engineering \\
  Atilim University \\
  Ankara, TR\\
  \texttt{sacip.toker@atilim.edu.tr} \\
}

\begin{document}
\maketitle

\begin{abstract}
The rise of Generative AI (GenAI) tools, such as ChatGPT, has transformed how students access and engage with information, raising questions about their impact on learning outcomes and retention. This study investigates how GenAI (ChatGPT), search engines (Google), and e-textbooks influence student performance across tasks of varying cognitive complexity, based on Bloom’s Taxonomy. Using a sample of 123 students, we examined performance in three tasks: [1] knowing and understanding, [2] applying, and [3] synthesizing, evaluating, and creating. Results indicate that ChatGPT and Google groups outperformed the control group in immediate assessments for lower-order cognitive tasks, benefiting from quick access to structured information. However, their advantage diminished over time, with retention test scores aligning with those of the e-textbook group. For higher-order cognitive tasks, no significant differences were observed among groups, with the control group demonstrating the highest retention. These findings suggest that while AI-driven tools facilitate immediate performance, they do not inherently reinforce long-term retention unless supported by structured learning strategies. The study highlights the need for balanced technology integration in education, ensuring that AI tools are paired with pedagogical approaches that promote deep cognitive engagement and knowledge retention.

\end{abstract}


\section{Introduction}
The integration of technology in education has transformed how students engage with learning tasks. Notably, Generative AI (GenAI) tools like ChatGPT have introduced new possibilities and challenges. These advanced AI systems offer the ability to synthesize information from various sources, enabling students to write essays and apply knowledge in new contexts with unprecedented ease. This represents a significant shift from previous reliance on search engines like Google, where students manually gathered and integrated information, to a more streamlined and automated process facilitated by GenAI.

Search engines are used to access a wide array of information in learning settings. They facilitate the retrieval of diverse sources, requiring students to critically analyze and synthesize data to complete assignments. This process is crucial for tasks that align with higher levels of Bloom's taxonomy, such as analysis, evaluation, and creation. While search engines provide the necessary factual information, the responsibility for integrating and synthesizing this information rests with the students, promoting deeper understanding and application.

The emergence of GenAI tools has streamlined this process, providing synthesized information and ready-to-use insights, which potentially reduces the cognitive load on students. While this capability can enhance efficiency, it raises important questions about the long-term implications for learning. Specifically, it is essential to understand whether the use of GenAI tools impacts student learning and whether their potential impacts vary depending on the complexity of the tasks students cope with. 

This study explores the role of different technologies—GenAI tools such as ChatGPT, search engines like Google, and traditional textbooks—in influencing students' achievement across learning tasks of varying complexities. By analyzing tasks that span the spectrum of Bloom's taxonomy, from knowledge recall to complex application and evaluation, this research seeks to determine the distinct impacts of these technologies on learning outcomes. Additionally, it examines whether the effectiveness of each technology varies with the complexity of tasks.

The insights gained from this study are critical for educators as they navigate the evolving educational landscape. Understanding the differential impacts of GenAI, search engines, and textbooks can inform the development of balanced instructional strategies that optimize the benefits of each tool while fostering essential cognitive skills. As the educational fields, including computer science education, continue to adapt to technological advancements, this research provides insights for informed decisions on integrating technology to support student learning and achievement.
  
\section{Background}
\subsection{E-textbooks and Digital Learning
}
E-textbooks have become crucial in higher education due to their convenience, searchability, and multimedia features \cite{woody2010books}. This shift aligns with the broader trend toward digital reading, which has transformed traditional reading habits into faster, more non-linear navigation \cite{mangen2008hypertext}. The widespread use of e-books, e-textbooks, and devices like Kindles and iPads in higher education has thus created a comprehensive digital reading environment \cite{koolen2012electronic}. Within this environment, students tend to prefer user-friendly resources that provide direct access to texts and yield scholarly results \cite{purdy2012first}.

As digital reading becomes more integrated into educational practices, advanced search systems have become essential to these platforms. These systems, which are highly rated by students for their ease of use, content quality, and connectivity \cite{purdy2012first}, now support enhanced learning through interactive environments. Therefore, understanding how students interact with and benefit from these digital resources is vital for maximizing their educational benefits.
\subsection{ 
Search Systems and Learning
}
Over the past two decades, information search systems have advanced from basic information retrieval tools to comprehensive full-text information systems. Researchers now view these search systems as rich online environments where users can learn and uncover new knowledge through interaction with online content\cite{hansen2016recent}. Consequently, searching and learning are regarded as interconnected within the field of information science. In recent years, the idea of designing search systems to enhance learning during the search process and create a rich educational environment has gained increasing recognition among researchers and practitioners.
Several studies have explored the relationship between search systems and learning. For instance, Lu and Hsiao \cite{lu2017personalized}) examined programming learners' information-seeking behaviors in online forums, revealing that advanced learners effectively refine queries and engage deeply with content while novices struggle. Roy et al. \cite{roy2020exploring} investigated the learning gains within search sessions, finding that participants with prior knowledge experienced significant learning gains by the end of their search sessions. This study highlighted the importance of prior knowledge in maximizing the educational benefits of search activities. Moraes et al. \cite{moraes2018contrasting} compared search-based learning with instructor-designed learning. They discovered that while watching lecture videos led to higher learning gains than searching alone, integrating search with lecture viewing significantly enhanced learning outcomes. Their results indicate that combining these approaches can be particularly effective.
\subsection{Learning through Search: A Taxonomical Approach
}
Bloom’s taxonomy is a hierarchical framework that outlines cognitive processes, suggesting that learning at a higher level depends on skills and knowledge attained at a lower level. Initially introduced by Bloom et al. (1956) and later revised by Anderson and a team of cognitive psychologists \cite{anderson2001taxonomy}, the updated version identifies six key categories of cognitive processes: Remember, Understand, Apply, Analyze, Evaluate, and Create. These categories vary in complexity, with 'Remember' being the simplest and 'Create' being the most complex.

Bloom’s taxonomy has been widely adopted by educators to set lesson objectives, design activities with varying levels of difficulty, and develop assessments. In computing education, it has been applied to evaluate engineering curricula \cite{dolog2016assessing}, create assessment items targeting different cognitive processes (e.g., \cite{khairuddin2008application}), measure student progress in programming courses (e.g., \cite{cabo2015quantifying}), understand the cognitive demands of programming (e.g., \cite{thompson2008bloom}), and define measurable learning outcomes (e.g., \cite{starr2008bloom}).

Vakkari suggested that Bloom's taxonomy can be applied to common search tasks involving querying to find topically relevant information or sources \cite{vakkari2016searching}. These tasks include formulating searches and evaluating results (source selection). Searchers require factual knowledge to choose search terms and identify relevant sources from result lists. They also utilize procedural knowledge to execute the search process. The cognitive processes involved in these tasks primarily include remembering—recognizing search terms from the task and results, recalling them from memory, and identifying relevant items in the result list. 
The taxonomy has been used to characterize learning requirements in search tasks (e.g., \cite{jansen2009using}, \cite{arguello2012task}). Jansen et al. explored whether the learning process has distinct information-searching characteristics \cite{jansen2009using}. Their laboratory experiment involving 72 participants and 426 search tasks revealed that middle-level cognitive processes such as applying and analyzing required the most searching effort. Surprisingly, the searching behaviors for the lowest (remembering and understanding) and highest (synthesizing, evaluating, and creating) learning categories were similar, indicating that searchers often use simple search expressions for higher-level information needs, primarily relying on internal knowledge for complex tasks. In a related line of inquiry, Arguello et al. examined how task complexity and the integration of vertical results (e.g., images, video, news) in aggregated search interfaces affect user interaction \cite{arguello2012task}. Their study with 29 subjects performing six search tasks of varying complexity showed that more complex tasks led to significantly more interaction and increased examination of vertical results. These studies collectively highlight the evolving role of search systems in learning, demonstrating that thoughtful design and user strategies can greatly enhance the educational potential of search activities.

\section{Present Study}
Several studies highlight the challenges search engines face in comprehending students' information needs due to their use of long, natural language phrases and ambiguous queries (e.g., \cite{bilal2000children},\cite{druin2010children}, \cite{jochmann2010children}). Despite recent developments to address this issue in information retrieval, formulating queries to access information has remained a challenge for students. This issue can be addressed using GenAI tools (e.g., \cite{zhang2024google}). 

GenAI tools are designed to process and understand natural language more effectively than traditional search algorithms. They can interpret complex, colloquial, and context-specific language, which aligns well with how children typically phrase their queries. By leveraging advanced natural language processing (NLP) capabilities, GenAI can more accurately decipher the intent behind children's questions, even when they are expressed in lengthy or vague terms. The emergence of GenAI tools such as ChatGPT has led learners to increasingly rely on these conversational agents (e.g., \cite{jalon2024chatgpt}). This transformation prompts critical inquiries about the role of AI tools in the learning process and their potential impact on memory and retention.
The rapid integration of GenAI tools in educational settings necessitates a deeper understanding of their impact on learning outcomes. While GenAI offers significant advantages in terms of efficiency and ease of information retrieval, it is crucial to investigate whether these benefits translate into improved learning, particularly across tasks of varying complexity, as outlined in Bloom’s taxonomy. Traditional search engines and textbooks have long been used to promote critical thinking and deeper cognitive engagement by requiring students to manually analyze, synthesize, and evaluate information. In contrast, GenAI tools streamline this process, potentially reducing the cognitive effort required by students.
This study aims to fill a gap in the existing literature by comparing the impacts of GenAI tools, search engines, and e-textbooks on student achievement. Specifically, it will examine how each technology influences learning across different cognitive levels—from basic knowledge recall to higher-order thinking skills such as evaluation and creation. Understanding these differences is essential for educators who must decide how to integrate these technologies into their teaching strategies effectively.
Furthermore, this research will explore whether the effectiveness of each technology varies with the complexity of the learning tasks. By doing so, it seeks to provide actionable insights that can help educators optimize the use of technological tools in fostering comprehensive learning experiences. As education continues to evolve with technological advancements, it is imperative to assess how these tools can be best utilized to support student learning and achievement in a balanced and effective manner. Specifically, we aim to answer the following research questions in this study.
\begin{itemize}
    \item Does the technology used (e-textbook, Google, ChatGPT groups vs. the control group) influence the achievement scores after completing three progressing tasks: knowing and understanding (Task 1), applying (Task 2), synthesizing, evaluating, and creating (Task 3)? 
    \item How does the technology used (e-textbook, Google, ChatGPT, or none) influence students’ retention of learning outcomes at different levels of Bloom’s Taxonomy? 

\end{itemize}

\section{Method}
\subsection{Participants}
152 students from an information systems department at a private university agreed to participate in this study. However, data from 24 students who did not complete all required tasks were excluded. Five students who failed to follow instructions accurately were also removed from the study. Consequently, the final sample consisted of 123 students.
Participants completed two pretests and two self-efficacy scales to measure their knowledge of data privacy, familiarity with ChatGPT, search self-efficacy, and search experiences. The average score for the data privacy test was 54.81 (SD = 27.10) out of 100, while the ChatGPT knowledge test averaged 71.29 (SD = 9.32) out of 100. The mean score for the search self-efficacy test was 49.52 (SD = 9.22) out of 70.
\subsection{Research Design}
This study employed an experimental pre-and post-test design with random assignment. This design included a control and three experimental groups, each exposed to different tools. Initially, participants were randomly assigned to one of four groups: control, e-textbook, Google, and ChatGPT access. The e-textbook group had access to a PDF version of the course textbook on data privacy, with no other resources allowed. The Google group used only the Google search engine, and the ChatGPT group relied solely on ChatGPT. The control group had no access to these tools during the experiment.
The experiment was conducted in a laboratory environment under the supervision of the course instructor and teaching assistants. Although students often use multiple resources simultaneously in real-world settings, this study restricted access to the assigned tools to ensure adherence to the protocol. Participants were continuously monitored, and only five students initially deviated from the protocol. To maintain internal validity, these students were reminded to use the designated tool and were subsequently excluded from the study.
Based on Bloom’s Revised Taxonomy of Educational Objectives \cite{krathwohl2002revision}, three tasks of varying complexity were designed for this study. Task 1 required lower-order thinking skills, such as remembering and understanding (see Table 1). Task 2 involved medium-order thinking skills, specifically applying knowledge (see Table 2). Task 3 demanded higher-order thinking skills, including analyzing, evaluating, and creating (see Table 3). All participants completed the tasks sequentially: Task 1, Task 2, and Task 3.

\begin{table}[h!]
\caption{Task 1 - Remembering and Understanding} 

\begin{tabular} {|p{16 cm}|}
\hline
Please answer the following questions and submit them as a Word file to the Moodle Course site.
\begin{itemize}
  \item What is the European Union (EU) General Data Protection Regulation (GDPR)? List the GDPR requirements that must be met by EU organizations when processing personal information including but not limited to collection recording organization structuring storing adaptation retrieval use dissemination combination and destruction of personal data.
  \item What types of information are protected by the GDPR?
  \item What are the main differences and similarities between privacy security and ethics?
  \item What do cookies mean? And why are they important for privacy?
\end{itemize}
\\ 
\hline
\end{tabular}
\end{table}

\begin{table}[h!]
\centering
\caption{Task 2 - Applying}
\begin{tabular} {|p{16 cm}|}
\hline
Suppose that you work for a company as a privacy and security expert with proficiency on legal and technical aspects of GDPR. The company’s management wants to make sure it has policies and procedures in place to protect the privacy of visitors to its website. You have been asked to provide advice to the company about what needs to be included in the policies to ensure that they comply with the EU GDPR. Your advice should address issues including but not limited to:
\begin{itemize}
  \item How much data should the company collect on visitors to its website and why that data? What information could it discover by tracking visitors’ activities on its website?
  \item What value would this information provide the company? What are the privacy problems raised by collecting such data?
  \item Should the company use cookies? What are the advantages of using cookies for both the company and its website visitors? What privacy issues do they create for the company?
  \item Should the company adopt an opt-in or opt-out model of informed consent?
\end{itemize}
\\ 
\hline
\end{tabular}
\end{table}

\begin{table}[h!]
\centering
\caption{Task 3 - Analyzing, Evaluating, and Creating}
\begin{tabular}{|p{16cm}|}
\hline
Examine the Optimum Customer Privacy Notice and evaluate whether it complies with the EU GDPR. Base your evaluation on the GDPR requirements, including but not limited to:
\begin{itemize}
  \item requiring organizations to protect PII (personally identifiable information)
  \item requiring organizations to allow users to access all their personal information without charge within one month
  \item requiring organizations to delete personal data (right to be forgotten)
  \item requiring organizations to ensure data portability so consumers are not locked into a particular service
  \item requiring organizations to guarantee the right to sue providers for damages or abuse of PII including class action lawsuits
  \item requiring organizations to have a data protection officer that reports to senior management;
  \item requiring explicit consent before collecting data (positive opt-in)
  \item requiring organizations to eliminate default opt-in processes
  \item requiring organizations to publish the rationale for data collection and the length of retention
  \item requiring organizations to report breaches and hacks within 72 hours
  \item requiring organizations to assume liability for data they share with partners or other firms
  \item requiring organizations to build privacy protections into all new systems (privacy by design)
  \item requiring organizations to anonymize data rather than targeting based on intimate personal profiles;
  \item requiring organizations to limit the collection of personal data to only that which is needed to support a task or a transaction and then deleting it shortly thereafter.
\end{itemize}
After the evaluation identify gaps in the privacy policy. What are your suggestions for improving the privacy policy (PR)? Please prepare a list of suggestions to improve the PR document.
\\ 
\hline
\end{tabular}
\end{table}

\subsection{Group Formation}

Participants were randomly assigned to one of four groups (control, e-textbook, Google, ChatGPT) at the start of the study. To ensure a balanced distribution of prior knowledge and experience across groups, a two-step cluster analysis was conducted. This clustering was based on factors such as search self-efficacy, prior knowledge of data privacy, and familiarity with AI tools (e.g., ChatGPT). This approach helped achieve a comparable distribution of experience levels across all experimental conditions.

The analysis resulted in four clusters:

\begin{itemize}
    \item Cluster 1 (24 participants) had very low composite scores across these factors.
    \item Cluster 2 (32 participants) had slightly higher but still low scores.
    \item Cluster 3 (62 participants) had medium scores.
    \item Cluster 4 (35 participants) had the highest scores.
\end{itemize}

To maintain balanced representation, participants from each cluster were evenly distributed across the four experimental groups, ensuring a diverse mix of experience and knowledge levels in each condition. Gender balance was also maintained during this process. The final gender distribution was 40 female students (32.52\%) and 83 male students (67.48\%). The control group consisted of 31 students (25.20\%), while the e-textbook, Google, and ChatGPT groups had 25 (20.33\%), 29 (23.58\%), and 38 (30.89\%) students, respectively. 

Overall, this method ensured that each group had a fair representation of students with varying levels of expertise, promoting the reliability of the study's findings.

\subsection{Procedure}

Groups completed three learning activities in the study. Each learning activity consisted of two components: (1) a task completed during the lab session, where students used their assigned tools (e-textbook, Google, ChatGPT, and none) to work through scenario-based exercises, and (2) a quiz administered in Moodle immediately after the task, featuring both multiple-choice and open-ended questions designed to assess individual comprehension without access to external tools.

For each task, one lecture/demonstration session (50 minutes) and one lab session (100 minutes) were allocated in the course. The lecture/demonstration session took place before the lab session and was used to review relevant course content, introduce key concepts, and demonstrate how they could be applied to similar tasks. In the lab sessions, students completed their assigned tasks with full access to their designated tools (e-textbook, Google, or ChatGPT) to support their engagement with the scenario-based exercises.

However, during the quiz phase, all students completed the Moodle-based assessment independently, without access to any external tools. This restriction ensured that the quizzes measured individual comprehension and retention of knowledge rather than immediate information retrieval from external resources.

While tasks required students to engage in recalling, comprehending, applying, analyzing, and creating using the provided resources, the quizzes independently evaluated their grasp of key concepts while also assessing the cognitive skills targeted in the corresponding task. For example, quizzes following Task 3 included items that measured students' ability to analyze, evaluate, and generate solutions in addition to recall and comprehension questions, ensuring alignment between tasks and assessment measures.

Three weeks after completing the final task, students took a post-test designed to assess long-term retention of key concepts. The post-test did not require students to replicate the original tasks but instead included multiple-choice and short-answer questions evaluating their understanding of the broader subject matter. While these questions were conceptually aligned with the original task assessments, they were reworded to minimize reliance on memorization. This approach ensured that the post-test measured students' ability to retain and apply knowledge beyond specific examples encountered in the initial tasks.

Figure 1 illustrates the overall research design and procedure followed in the study.

\begin{figure}[h!]
\centering
\includegraphics{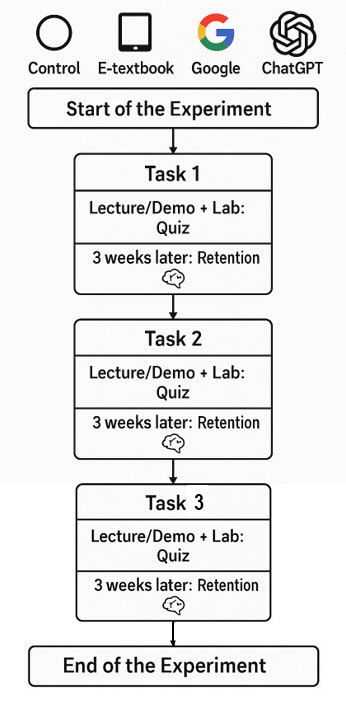}
\caption{Overall research design and data collection}
\end{figure}

\subsection{Data Analysis and Measures}
For the first research question, we collected assessment test scores for three progressive tasks, which served as the dependent variables. The independent variable was the group category, consisting of four groups: three experimental groups and one control group. MANOVA analyses were performed \cite{field2013discovering}. After a significant result, we followed up univariate ANOVAs for each task assessment. We checked the assumptions for both MANOVA and univariate ANOVA. Post-hoc tests, such as Scheffe or Dunnett's C, were used to determine which specific group caused the significant results. For the second research question, we used repeated-measures ANOVA. This was appropriate because assessment and retention test scores were measured on two different occasions for the same groups \cite{field2013discovering}.
We ensured the content validity of the assessments by aligning them with the goals of the tasks and the specifics of the privacy topics. For each task, specific goals were identified based on their complexity and matched in detail with the relevant privacy topics. The instructor and teaching assistant reviewed the assessments to confirm their content validity. 

\section{Results}
\subsection{Results for Research Question 1}
Table 4 illustrates descriptive statistics about the assessment tests for three tasks across four study groups: Control, E-textbook, Google, and ChatGPT. 

\begin{table}[htbp]
    \centering
    \caption{Descriptive information about the assessment tests for all tasks and study groups}
    \begin{tabular}{llcc}
        \toprule
        \textbf{Task} & \textbf{Group} & \textbf{M} & \textbf{SD} \\
        \midrule
        \multirow{5}{*}{\textbf{Task 1}} 
            & Control & 60,2 & 17,6 \\
            & E-textbook & 66,0 & 21,4 \\
            & Google & 79,1 & 17,4 \\
            & ChatGPT & 82,6 & 17,8 \\
            & \textbf{Total} & \textbf{72,8} & \textbf{20,5} \\
        \midrule
        \multirow{5}{*}{\textbf{Task 2}} 
            & Control & 47,7 & 19,8 \\
            & E-textbook & 60,8 & 22,9 \\
            & Google & 66,6 & 13,7 \\
            & ChatGPT & 67,9 & 20,0 \\
            & \textbf{Total} & \textbf{61,1} & \textbf{20,8} \\
        \midrule
        \multirow{5}{*}{\textbf{Task 3}} 
            & Control & 46,0 & 14,7 \\
            & E-textbook & 42,7 & 12,1 \\
            & Google & 41,5 & 14,8 \\
            & ChatGPT & 41,2 & 14,8 \\
            & \textbf{Total} & \textbf{42,8} & \textbf{14,2} \\
        \bottomrule
    \end{tabular}
\end{table}

There were significant differences in task-level assessment scores based on the groups. The multivariate analysis indicated a significant impact of groups on the three task-level assessment scores, with Wilk’s \textlambda = .677, F(9, 284.90) = 5.513, p < .001, and partial η² = .122, explaining 12.2\% of the total variance. Univariate ANOVA findings were examined. 

\subsubsection{Task 1 Assessment Scores
}
The univariate ANOVA results for Task 1 showed significant differences between groups, F(3, 119) = 10.650, p < .001, partial η² = .212, explaining 21.2\% of the variance. The ChatGPT group outperformed the E-textbook and control groups significantly, with mean differences of 16.60 and 22.39, respectively (p < .01). The Google group also scored significantly higher than the control group, with a mean difference of 18.90 (p < .01). The ChatGPT group demonstrated the highest performance in Task 1, as illustrated in Figure 2.

\begin{figure}[h!]
\centering
\includegraphics[width=\columnwidth]{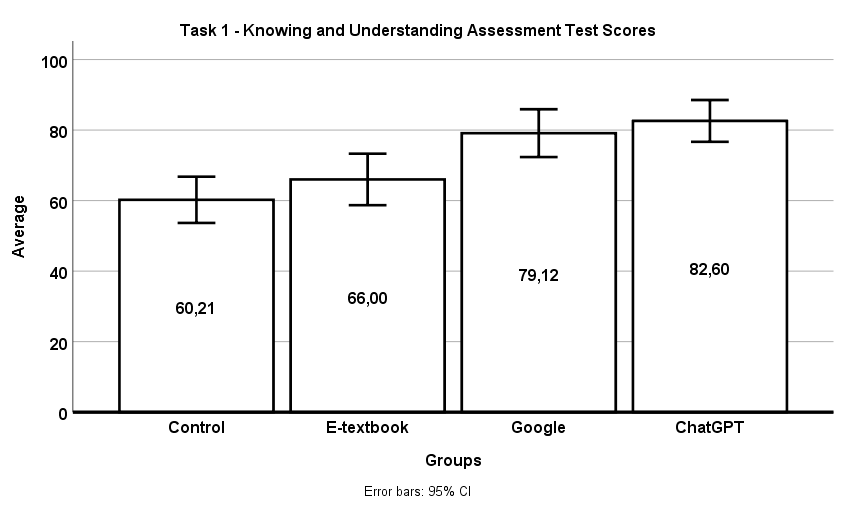}
\caption{Task 1 Assessment Scores}
\end{figure}

\subsubsection{Task 2 Assessment Scores
}

For Task 2, the univariate ANOVA results also showed significant differences between groups, F(3, 119) = 7.279, p < .001, partial η² = .155, explaining 15.5\% of the variance. The control group scored significantly lower than both the Google and ChatGPT groups, with mean differences of 18.81 and 20.15, respectively (p < .01). The ChatGPT group again showed the best performance, while the control group had the lowest scores in Task 2 (see Figure 3).  

\begin{figure}[h!]
\centering
\includegraphics[width=\columnwidth]{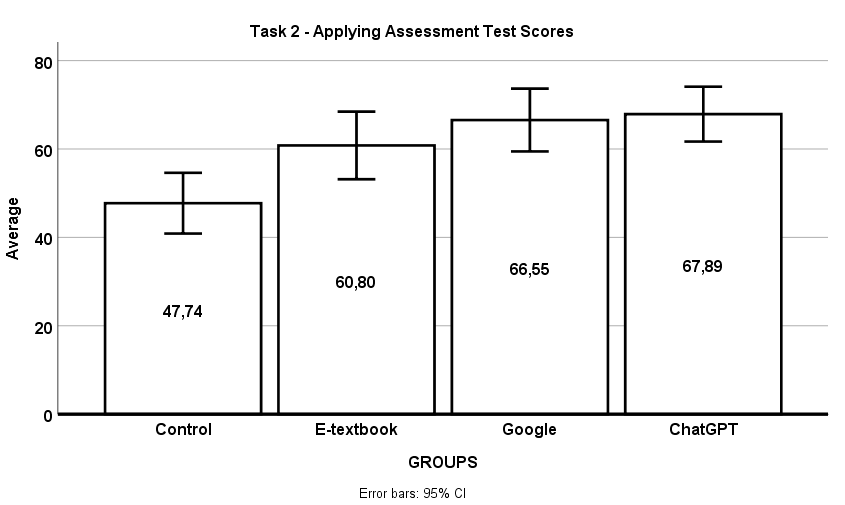}
\caption{Task 2 Assessment Scores}
\end{figure}

\subsubsection{Task 3 Assessment Scores
}
The univariate ANOVA results for Task 3 did not show significant differences between groups, F(3, 119) = .768, p = .514. The results indicated that the control group had the best performance, while the E-textbook, Google, and ChatGPT groups had lower scores compared to the control group (see Figure 4).

\begin{figure}[h!]
\centering
\includegraphics[width=\columnwidth]{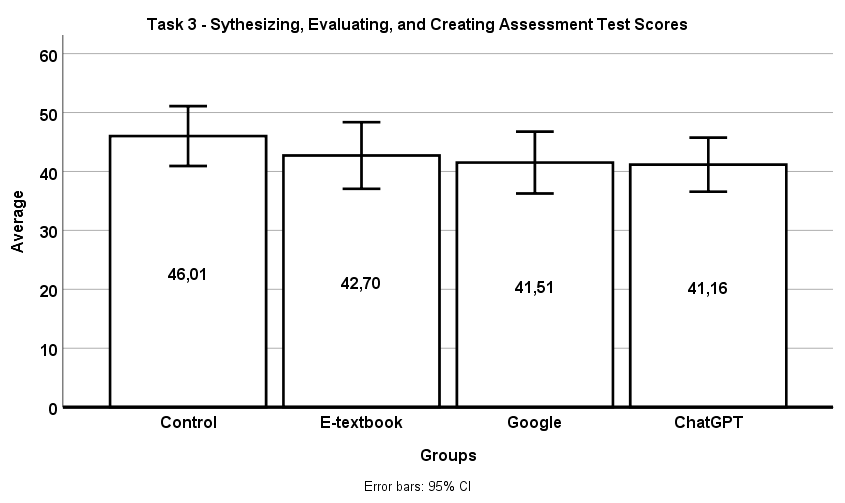}
\caption{Task 3 Assessment Scores}
\end{figure}

\subsection{Results for Research Question 2
}
The statistical analysis results of the study reveal important insights into the performance and retention of different groups across three tasks.

\begin{table}[htbp]
    \centering
    \caption{Descriptive information about retention tests}
    \begin{tabular}{llcc}
        \toprule
        \textbf{Task} & \textbf{Group} & \textbf{Mean} & \textbf{SD} \\
        \midrule
        \multirow{4}{*}{\textbf{Task 1}} & Control & 59.3 & 22.06 \\
        & E-textbook & 67.6 & 24.94 \\
        & Google & 67.8 & 25.4 \\
        & ChatGPT & 65.5 & 22.1 \\
        \midrule
        \multirow{4}{*}{\textbf{Task 2}} & Control & 45.9 & 20.0 \\
        & E-textbook & 61.3 & 22.1 \\
        & Google & 63.1 & 20.1 \\
        & ChatGPT & 64.5 & 21.1 \\
        \midrule
        \multirow{4}{*}{\textbf{Task 3}} & Control & 44.2 & 16.3 \\
        & E-textbook & 41.9 & 14.9 \\
        & Google & 40.6 & 13.8 \\
        & ChatGPT & 39.9 & 15.1 \\
        \bottomrule
    \end{tabular}
\end{table}

\subsubsection{Task 1 Results}
Descriptive statistics for retention tests are provided in Table 5.
The interaction effect between assessment and retention scores was significant (Wilk’s \textlambda = .885, F(3, 119) = 5.175, p < .01, partial \texteta = .115), explaining 11.5\% of the variance (for Task 1). The significant interaction indicates that changes in scores between the assessment and retention tests differed across groups. Specifically, the ChatGPT group experienced a notable drop in retention scores, while the Google group also saw a decline, albeit less pronounced. The control and E-textbook groups exhibited smaller fluctuations.
These findings suggest that while the ChatGPT and Google groups performed highly immediately after the task, they struggled to retain the learned information over time. In contrast, the control and e-textbook groups had more stable performance and retention.

\subsubsection{Task 2 Results}


None of the effects produced significant results. Although there were decreases in performance for the ChatGPT and Google groups and increases for the E-textbook and control groups, these changes were not statistically significant (see Table 5 for Task 2). This suggests that while there were variations in performance and retention, they were not substantial enough to be deemed significant.  

\subsubsection{Task 3 Results}


None of the effects produced significant results. There were decreases in performance across all groups from the assessment to the retention tests; however, these changes were not significant (see Table 5 for Task 3).

\section{Discussion}
The present study investigated the impact of different technologies — GenAI tools like ChatGPT, search engines like Google, and traditional e-textbooks—on students' learning outcomes across tasks of varying complexity, based on Bloom's taxonomy. The findings provide critical insights into how these technologies influence immediate performance and long-term retention.

\subsection{Impact on Immediate Performance}
The study found significant differences in performance across different technologies in tasks involving knowing, understanding, and applying (i.e., Task 1 and Task 2). Both the ChatGPT and Google groups outperformed the control group in these tasks, aligning with Vakkari’s findings on the relationship between Bloom’s taxonomy and search tasks \cite{vakkari2016searching}. This advantage was particularly evident in tasks focused on memorization and comprehension, where quick access to structured information plays a key role in performance.

A likely explanation for this pattern is that students using ChatGPT and Google were able to efficiently retrieve structured, relevant responses that directly aligned with the task requirements. This reduced cognitive effort in searching for and organizing information, allowing them to quickly process and reproduce factual knowledge. However, this form of engagement primarily supports surface-level learning, as it does not necessarily require students to critically evaluate, integrate, or generate new ideas from the retrieved content.

This distinction may explain why the performance advantage of ChatGPT and Google diminished in tasks requiring higher-order cognitive skills (e.g., Task 3). As tasks became more complex, requiring students to analyze, evaluate, or create new content, the ability to retrieve well-structured responses was no longer sufficient. These tasks demand active cognitive engagement beyond factual recall, which may not have been fully supported by the way students interacted with AI tools and search engines.

\subsection{Impact on Retention}
While ChatGPT and Google provided advantages in immediate performance, these benefits did not necessarily translate into enhanced long-term retention. In the retention tests conducted three weeks after the knowing and understanding tasks, the performance of the ChatGPT and Google groups declined and aligned with that of the e-textbook group. This reduction over time is consistent with natural forgetting processes and suggests that while these tools facilitate efficient access to information, they do not inherently strengthen memory pathways to ensure sustained retention.

For the applying task, although there was a decrease in performance from the initial assessment to the retention test, the change was not statistically significant. This suggests that while ChatGPT and Google facilitated immediate task completion, their use did not lead to notably superior retention compared to other tools. As tasks increased in complexity, the differences in retention across groups became less pronounced, indicating that no single tool provided a sustained advantage in long-term knowledge retention.

In the synthesizing, evaluating, and creating tasks, the control group demonstrated the highest retention scores. The decline in scores from assessment to retention tests was observed across all groups but was not statistically significant. Interestingly, the e-textbook group exhibited a smaller decrease, suggesting that the act of reading and comprehending textual information might have reinforced retention more effectively than retrieving pre-structured responses. This observation aligns with Purdy’s findings \cite{purdy2012first} on the effectiveness of e-textbooks in fostering deeper engagement with scholarly content.

The relatively lower retention in the GenAI and Google groups may be due to the manner in which students interacted with these tools. While these technologies provide quick access to synthesized information, they do not inherently reinforce knowledge integration unless paired with learning strategies that actively engage memory consolidation mechanisms. GenAI tools, including ChatGPT, sometimes generated responses that were too generic or lacked depth, limiting their effectiveness in supporting retention for complex tasks. Similarly, effective use of Google required deeper query formulation and critical filtering of information, which may have introduced variability in retention outcomes \cite{lu2017personalized}.

Ultimately, ensuring long-term retention requires instructional approaches that extend beyond initial task completion. Since human memory is susceptible to natural forgetting \cite{baddeley1997human}, additional reinforcement activities—such as reflective exercises \cite{clark2018assessing}, spaced repetition \cite{karpicke2011spaced}, and active recall techniques—may be necessary to solidify learning over time, regardless of the technology used. Future research should explore how structured learning interventions can be integrated with GenAI and search tools to maximize both immediate learning and long-term retention.

\subsection{Implications for Educational Practice}
This study suggests several implications for higher education. First, while GenAI tools like ChatGPT offer significant potential due to their ease of use and high-quality responses, they may not entirely replace the need for search engines or traditional learning resources. Educators should consider designing assignments that move beyond lower and medium-level cognitive tasks to higher-order skills, where the benefits of GenAI are complemented by deeper cognitive engagement.

\subsection{Limitations and Future Research Directions
}
This study has several limitations that should be considered when interpreting the findings.

First, the sequential nature of the tasks may have influenced group performance over time. Participants completed tasks in a fixed order, which could have introduced cumulative advantages or disadvantages, particularly for those using AI tools or search engines. This ordering may have impacted learning effects, making it difficult to isolate the influence of individual tasks. Future research should explore alternative study designs that counterbalance task order to mitigate potential sequencing effects.

Second, the sample size and participant demographics may limit the generalizability of the findings. The study was conducted with 123 students from a single institution, which, while providing valuable insights, may not fully capture the diverse ways students engage with learning technologies across different educational contexts. Expanding the research to include larger and more diverse participant pools would help validate and extend these findings.

Finally, the study was conducted in a controlled environment where participants were restricted to using only their assigned tools. However, in real-world academic settings, students frequently combine multiple resources, including AI tools, search engines, and textbooks, to complete learning tasks. Future studies should investigate how mixed-resource learning environments influence knowledge retention, cognitive engagement, and long-term skill development to better reflect authentic learning behaviors.

Future research should explore the relationship between prompting strategies and learning outcomes across different task complexities. Investigating how GenAI and search engine technologies can be integrated with instructional strategies to enhance long-term retention and skill acquisition is also essential. Understanding these dynamics will help educators leverage the strengths of each technology, fostering comprehensive learning experiences that support both immediate understanding and long-term retention.

While GenAI tools like ChatGPT provide substantial benefits for immediate learning and information retrieval, their limitations in supporting long-term retention and higher-order cognitive skills highlight the importance of a balanced instructional approach. By integrating these tools with appropriate instructional strategies, educators can create enriched learning environments that support students across all cognitive levels.

\bibliographystyle{unsrt}  
\bibliography{references}

\end{document}